\documentclass[conference]{IEEEtran}

\usepackage[nolist]{acronym} 
\usepackage{algorithm,algpseudocode}
\usepackage{pgfplots}
\usepgfplotslibrary{groupplots}
\usepackage{graphicx}
\pgfplotsset{compat=newest}
\usepackage[]{nohyperref}  
\usepackage{units}
\usepackage{amsmath, amsbsy, amssymb}
\usepackage{tikzscale}
\usepackage{color}
\usepackage{epigraph}
\usepackage{circuitikz}
\usepackage[group-separator={,}]{siunitx}
\usetikzlibrary{matrix}
\usetikzlibrary{positioning}
\usepackage{bbm}

\definecolor{mittelblau}{RGB}{0, 126, 198}
\definecolor{violettblau}{cmyk}{0.9, 0.6, 0, 0}
\definecolor{rot}{RGB}{238, 28 35}
\definecolor{apfelgruen}{RGB}{140, 198, 62}
\definecolor{gelb}{RGB}{1, 221, 0}
\definecolor{orange}{RGB}{244, 111, 33}
\definecolor{pink}{RGB}{237, 0, 140}
\definecolor{lila}{RGB}{128, 10, 145}
\definecolor{hellgrau}{RGB}{224, 224, 224}
\definecolor{mittelgrau}{RGB}{128, 128, 128}
\definecolor{dunkelgrau}{RGB}{80,80,80}
\definecolor{anthrazit}{RGB}{19, 31, 31}

\usetikzlibrary{shapes,arrows}

\IEEEoverridecommandlockouts

\begin{document}

\title{Avoiding Burst-like Error Patterns in Windowed Decoding of Spatially Coupled LDPC Codes}

\author{\IEEEauthorblockN{ Kevin Klaiber\IEEEauthorrefmark{1},  Sebastian Cammerer\IEEEauthorrefmark{1}, Laurent Schmalen\IEEEauthorrefmark{2}  and Stephan ten Brink\IEEEauthorrefmark{1}}

\IEEEauthorblockA{
\IEEEauthorrefmark{1} Institute of Telecommunications, Pfaffenwaldring 47, University of  Stuttgart, 70659 Stuttgart, Germany 
\\
\IEEEauthorrefmark{2}Nokia Bell Labs, Lorenzstr. 10, 70435 Stuttgart, Germany, 
}

}

\maketitle

\renewcommand{\vec}[1]{\mathbf{#1}}
\newcommand{\vecs}[1]{\boldsymbol{#1}}

\newcommand{\av}{\vec{a}}
\newcommand{\bv}{\vec{b}}
\newcommand{\cv}{\vec{c}}
\newcommand{\dv}{\vec{d}}
\newcommand{\ev}{\vec{e}}
\newcommand{\fv}{\vec{f}}
\newcommand{\gv}{\vec{g}}
\newcommand{\hv}{\vec{h}}
\newcommand{\iv}{\vec{i}}
\newcommand{\jv}{\vec{j}}
\newcommand{\kv}{\vec{k}}
\newcommand{\lv}{\vec{l}}
\newcommand{\mv}{\vec{m}}
\newcommand{\nv}{\vec{n}}
\newcommand{\ov}{\vec{o}}
\newcommand{\pv}{\vec{p}}
\newcommand{\qv}{\vec{q}}
\newcommand{\rv}{\vec{r}}
\newcommand{\sv}{\vec{s}}
\newcommand{\tv}{\vec{t}}
\newcommand{\uv}{\vec{u}}
\newcommand{\vv}{\vec{v}}
\newcommand{\wv}{\vec{w}}
\newcommand{\xv}{\vec{x}}
\newcommand{\yv}{\vec{y}}
\newcommand{\zv}{\vec{z}}
\newcommand{\zerov}{\vec{0}}
\newcommand{\onev}{\vec{1}}
\newcommand{\alphav}{\vecs{\alpha}}
\newcommand{\betav}{\vecs{\beta}}
\newcommand{\gammav}{\vecs{\gamma}}
\newcommand{\lambdav}{\vecs{\lambda}}
\newcommand{\omegav}{\vecs{\omega}}
\newcommand{\sigmav}{\vecs{\sigma}}
\newcommand{\tauv}{\vecs{\tau}}

\newcommand{\Am}{\vec{A}}
\newcommand{\Bm}{\vec{B}}
\newcommand{\Cm}{\vec{C}}
\newcommand{\Dm}{\vec{D}}
\newcommand{\Em}{\vec{E}}
\newcommand{\Fm}{\vec{F}}
\newcommand{\Gm}{\vec{G}}
\newcommand{\Hm}{\vec{H}}
\newcommand{\Id}{\vec{I}}
\newcommand{\Jm}{\vec{J}}
\newcommand{\Km}{\vec{K}}
\newcommand{\Lm}{\vec{L}}
\newcommand{\Mm}{\vec{M}}
\newcommand{\Nm}{\vec{N}}
\newcommand{\Om}{\vec{O}}
\newcommand{\Pm}{\vec{P}}
\newcommand{\Qm}{\vec{Q}}
\newcommand{\Rm}{\vec{R}}
\newcommand{\Sm}{\vec{S}}
\newcommand{\Tm}{\vec{T}}
\newcommand{\Um}{\vec{U}}
\newcommand{\Vm}{\vec{V}}
\newcommand{\Wm}{\vec{W}}
\newcommand{\Xm}{\vec{X}}
\newcommand{\Ym}{\vec{Y}}
\newcommand{\Zm}{\vec{Z}}
\newcommand{\Lambdam}{\vecs{\Lambda}}
\newcommand{\Pim}{\vecs{\Pi}}

\newcommand{\Ac}{{\cal A}}
\newcommand{\Bc}{{\cal B}}
\newcommand{\Cc}{{\cal C}}
\newcommand{\Dc}{{\cal D}}
\newcommand{\Ec}{{\cal E}}
\newcommand{\Fc}{{\cal F}}
\newcommand{\Gc}{{\cal G}}
\newcommand{\Hc}{{\cal H}}
\newcommand{\Ic}{{\cal I}}
\newcommand{\Jc}{{\cal J}}
\newcommand{\Kc}{{\cal K}}
\newcommand{\Lc}{{\cal L}}
\newcommand{\Mc}{{\cal M}}
\newcommand{\Nc}{{\cal N}}
\newcommand{\Oc}{{\cal O}}
\newcommand{\Pc}{{\cal P}}
\newcommand{\Qc}{{\cal Q}}
\newcommand{\Rc}{{\cal R}}
\newcommand{\Sc}{{\cal S}}
\newcommand{\Tc}{{\cal T}}
\newcommand{\Uc}{{\cal U}}
\newcommand{\Wc}{{\cal W}}
\newcommand{\Vc}{{\cal V}}
\newcommand{\Xc}{{\cal X}}
\newcommand{\Yc}{{\cal Y}}
\newcommand{\Zc}{{\cal Z}}

\newcommand{\CN}{\Cc\Nc}

\newcommand{\CC}{\mathbb{C}}
\newcommand{\MM}{\mathbb{M}}
\newcommand{\NN}{\mathbb{N}}
\newcommand{\RR}{\mathbb{R}}

\newcommand{\htp}{^{\mathsf{H}}}
\newcommand{\tp}{^{\mathsf{T}}}

\newcommand{\LB}{\left(}
\newcommand{\RB}{\right)}
\newcommand{\LP}{\left\{}
\newcommand{\RP}{\right\}}
\newcommand{\LSB}{\left[}
\newcommand{\RSB}{\right]}

\renewcommand{\ln}[1]{\mathop{\mathrm{ln}}\LB #1\RB}
\newcommand\norm[1]{\left\lVert#1\right\rVert}
\newcommand{\cs}[1]{\mathop{\mathrm{cs}}\LSB #1\RSB}

\newcommand{\EE}{{\mathbb{E}}}
\newcommand{\Expect}[2]{\EE_{#1}\LSB #2\RSB}

\newtheorem{definition}{Definition}[section]
\newtheorem{remark}{Remark}

\begin{acronym}
 \acro{ADC}{analog-to-digital converter}
 \acro{AGC}{automatic gain control}
 \acro{ASIC}{application-specific integrated circuit}
 \acro{AWGN}{additive white Gaussian noise}
 \acro{BER}{bit error rate}
 \acro{BICM}{bit interleaved coded modulation}
 \acro{BLER}{block error rate}
 \acro{CFO}{carrier frequency offset}
 \acro{DL}{deep learning}
 \acro{DQPSK}{differential quadrature phase-shift keying}
 \acro{ECC}{error correcting code}
 \acro{FPGA}{field programmable gate array}
 \acro{GNR}{GNU Radio}
 \acro{GPU}{graphic processing unit}
 \acro{ISI}{inter-symbol interference}
 \acro{LOS}{line-of-sight}
 \acro{MIMO}{multiple-input multiple-output}
 \acro{ML}{machine learning}
 \acro{MLP}{multilayer perceptron}
 \acro{MSE}{mean squared error}
 \acro{NN}{neural network}
 \acro{PLL}{phase-locked loop}
 \acro{ppm}{parts per million}
 \acro{PSK}{phase-shif keying}
 \acro{PFB}{polyphase filterbank}
 \acro{QAM}{quadrature amplitude modulation}
 \acro{ReLU}{rectified linear unit}
 \acro{RNN}{recurrent neural network}
 \acro{RRC}{root-raised cosine}
 \acro{RTN}{radio transformer network}
 \acro{SDR}{software-defined radio}
 \acro{SFO}{sampling frequency offset}
 \acro{SGD}{stochastic gradient descent}
 \acro{SNR}{signal-to-noise ratio}	
 \acro{TDL}{tapped delay line}
 \acro{OFDM}{orthogonal frequency division multiplex}
 \acro{IFFT}{inverse fast Fourier transform}
 \acro{FFT}{fast Fourier transform}
 \acro{IFT}{inverse Fourier transform}
 \acro{FT}{Fourier transform}
 \acro{IDFT}{inverse discrete Fourier-transform}
 \acro{DFT}{discrete Fourier-transform}
 \acro{CP}{cyclic prefix}
 \acro{MMSE}{minimum mean squared error}
 \acro{QPSK}{quadrature phase-shift keying}
 \acro{BP}{belief propagation}
 \acro{SC}{spatial coupling}
 \acro{LDPC}{low-density parity-check}
 \acro{SC-LDPC}{spatially coupled low-density parity-check}
 \acro{DE}{density evolution}
 \acro{MAP}{maximum a posteriori}
 \acro{VN}{variable node}
 \acro{CN}{check node}
 \acro{LLR}{log likelihood ratio}
\end{acronym}

\begin{abstract}

In this work, we analyze efficient window shift schemes for windowed decoding of \ac{SC-LDPC} codes, which is known to yield close-to-optimal decoding results when compared to \emph{full} \ac{BP} decoding. However, a drawback of windowed decoding is that either a significant amount of window updates are required leading to unnecessary high decoding complexity or the decoder suffers from sporadic \emph{burst-like} error patterns, causing a \emph{decoder stall}. To tackle this effect and, thus, to reduce the average decoding complexity, the basic idea is to enable adaptive window shifts based on a \ac{BER} prediction, which reduces the amount of unnecessary updates.
As the decoder stall does not occur in analytical investigations such as the \ac{DE}, we examine different schemes on a fixed test-set and exhaustive monte-carlo simulations based on our \ac{GPU} simulation framework.
As a result, we can reduce the average decoding complexity of the \emph{naive} windowed decoder while improving the \ac{BER} performance when compared to a non-adaptive windowed decoding scheme.
Furthermore, we show that a foresightful stall \emph{prediction} does not significantly outperform a retrospective stall \emph{detection} which is much easier to implement in practice.
\end{abstract}

\acresetall

\section{Introduction}

The general concept of \ac{SC} of codes, i.e., to locally connect multiple versions of a same underlying block code, has been shown to result in powerful code constructions, with excellent \ac{BER} performance \cite{lentmaier2010iterative} and a universal behavior with respect to the channel front-end \cite{Coupl11BMS,Schmalentenbrink}. This superior performance has been analytically shown in \cite{Coupl11BMS,yedla2012simple} and it turns out that, for carefully chosen coupling and code parameters, the \ac{BP} decoding threshold converges towards the \ac{MAP} decoding threshold of the underlying block code. This effect is known as \emph{threshold saturation} \cite{Coupl11BMS}. 
However, in practice, the price to pay is typically a high number of \ac{BP} decoding iterations when decoded with the naive version (\emph{block}-based) of the \ac{BP} decoder leading to high decoding complexity due to many unnecessary node updates.

To overcome this limitation of \ac{SC-LDPC} codes, a windowed decoding scheme has been proposed in \cite{iyengar2012windowed} and further analyzed in \cite{iyengar2013windowed,hassan2017non}.
It turns out that windowed decoding does not significantly degrade the decoding thresholds nor the \ac{BER} performance for carefully chosen decoder parameters.
Although remarkable decoding thresholds are analytically achieved for the windowed decoder, practically choosing these decoder parameters, such as the number of iterations per window shift and the window size, is a non-trivial task and provides more degrees of freedom than in the \emph{conventional, block-based} decoder.
The windowed decoder uses the fact that the \ac{BER} per spatial position converges in a \emph{wavelike} manner, i.e., subsequent blocks can only be decoded if the previous blocks have been successfully decoded. Therefore, it is sufficient to only update nodes within a few spatial positions and shift the active decoding window whenever a certain block is successfully decoded or a maximum number of iterations reached. Contrary to full \ac{BP} decoding, windowed decoding requires knowledge about the \emph{active} positions during decoding or, in other words, the decoder needs to track the \emph{decoding wave}.

In a straightforward implementation of windowed decoding, the window is sometimes shifted although a spatial position is still erroneous and, thus, decoding of all following blocks inherently fails, i.e., decoding is stuck.
In this work, we focus on \ac{SC-LDPC} codes, where this effect has been first reported in \cite{Schmalen2016Window}, however, similar observations have been later reported for braided codes in \cite{zhu2018braided}.
If not further analyzed, this effect simply shows up as an increased \ac{BER} in the \ac{SNR} range above the \ac{BP} threshold of the underlying block code, leading to a shifted waterfall region of the \ac{SC-LDPC} code. However, when carefully looking at the error distributions, this effect causes a \emph{burst}-like error distribution only in several decoded codewords which are only partly decoded.

We propose and compare adaptive windowed decoding schemes with respect to the window position and number of iterations to avoid burst-like errors.
Besides the exploration of adaptive window shift schemes, the main objective of this work is to examine whether it is possible to predict the occurrence of decoder stalls.   
This could lead to a further reduction of decoding complexity due to omitting unnecessary interventions to avoid a decoder stall. 
An empirical study of the problem seems promising, as the effect does not show up in the conventional threshold analysis.
To enable a more systematic analysis of the problem, we create a test-set consisting of noisy codewords causing decoder stalls for the \emph{naive} windowed decoder.
Finally, we use our test-set to answer the question whether a foresightful stall \emph{prediction} does significantly outperform a retrospective stall \emph{detection}.

\section{SC-LDPC and Windowed Decoding}
\label{sec:intro_sc}
\vspace*{-0.05cm}
To clarify notation, we provide a short \ac{SC-LDPC} introduction, for further details we refer to \cite{lentmaier2010iterative,Coupl11BMS}.
SC-LDPC codes can be seen as \ac{LDPC} codes that have a superimposed convolutional structure. The unit-memory \ac{SC-LDPC} code we consider in this work has a block-type parity-check matrix $\Hm_\text{sc}$ with matrix $\Hm_0$ in blocks indexed by positions $(j,j)$ and matrix $\Hm_1$ in blocks indexed by position $(j+1,j)$ for $j \in(1,L)$ and zero matrices in all other positions.
The sparse sub-matrices $\Hm_i$ of the \ac{SC-LDPC} parity-check matrix $\Hm_\text{sc}$ have size $\dim \Hm_i = m \times n$.
Assuming terminated SC-LDPC codes, the overall code length can be adapted by the replication factor $L$, whereby the overall block length amounts to $Ln$.
In this work, we use the same SC-LDPC code proposed in \cite{Schmalen2016Window} and, thus, reference the interested reader to \cite{Schmalen2016Window} for further details.
The most important characteristics of this code are the code rate $R$ $\approx$ 0.8 and degree distribution $d_\text{v} = 5$, $d_\text{c} = 25$. Furthermore, the code is non-uniformly coupled \cite{schmalen2017nonuniform} and optimized for a good threshold. Besides it has unit memory $\mu = 1$ and the two sub-matrices $\Hm_0$ and $\Hm_1$ are of size $\dim  \Hm = 960 \times 4800$. 
\vspace*{-0.1cm}
\subsection{Windowed Decoding}
\vspace*{-0.1cm}
If properly terminated, SC-LDPC codes can be decoded by the conventional \ac{BP} algorithm based on $\Hm_\text{sc}$, in the following referred to as \emph{full BP decoding}. However, in the context of large replication factors $L$ or streaming based data transmission, this procedure is not feasible due to decoding complexity and undesirable high latency. Both problems can be solved by introducing a windowed decoder \cite{iyengar2013windowed}, i.e., the message updates are only conducted within a certain window of size $w$. This \emph{decoding window} can now be shifted to the next position after having performed $I$ iterations inside the window while keeping the messages after each window shift. The number of windows $N_\text{w}$ denotes the required shifts to decode all spatial positions $L$. 
 
We define the average computational complexity $\overline{C}$ through the number of iterations per window ${I}_i$ and the window size $w_i$ of window $i$ as
$$\overline{C}=\frac{1}{N_\text{w}} \sum_{i=1}^{N_\text{w}} {I}_i \cdot w_{i}.$$
The decoding window can be defined either from \ac{VN} or \ac{CN} perspective. Throughout this work we opt for the \ac{CN} perspective, leading to an underlying parity-check matrix 
\begin{equation}
\Hm_{w}=
\begin{pmatrix}
\Hm_{\mu}		& \dots		& \Hm_1 	     & \Hm_0  	\\
&	\Hm_{\mu} & \dots		& \Hm_1 		 & \Hm_0	\\
&& \ddots	& \ddots	& \ddots 	 & \ddots			\\		
&&&	\Hm_{\mu} & \dots		& \Hm_1 		 & \Hm_0	\\
\end{pmatrix}_{w m \times (w + \mu)n \hspace*{-1cm}} 
\label{equ:generalCND}
\end{equation}
used within the windowed decoder.
Further, $p_\text{win} \in (1,N_{\text{win}})$\footnote{With abuse of some notation, we assume blocks outside $(1,L)$ are \emph{virtual} positions, initialized with known values. Thus, $N_{\text{win}} > L$, e.g. for the adaptive iteration decoder and fixed $w$ there are $N_{\text{win}}=L+ w +\mu -1$ windows.}  denotes the current window position, i.e, blocks in $(p_\text{win}, p_\text{win} + w + \mu -1)$ are active. In the following, we assume $\mu=1$.

\subsection{Decoder stall}

\begin{figure}
	\centering
	\resizebox{\columnwidth}{!}{
	\begin{tikzpicture}

\usepgfplotslibrary{colormaps}

\pgfplotsset{compat=1.12}
\tikzset{ellC/.style={/utils/exec={\pgfplotscolormapdefinemappedcolor{#1}},%
    draw=mapped color!80!mittelblau, fill=mapped color!80!white}}
colormap={slategraywhite}{rgb255=(112,128,144) rgb255=(255,159,101)},
\pgfplotsset{
	height=5.15cm,width=1.05\columnwidth,
	xmin=0.99, xmax=9.01,
        	ymin=3e-4,
	ymax=0.06,
	xtick distance=1,
	xlabel={Spatial position $i$ inside decoding window $w_\text{dec}$},
	ylabel={P_{\text{e,post-dec},i}},
	grid=both,
	ymode=log,
	legend style={legend pos=north west,font=\small}}

\pgfplotsset{%
    colormap={customcolor}{rgb255=(0, 126, 198) rgb255=(238, 28, 35)}
}%

\begin{axis}[%
name=plot1,
xlabel={Spatial position within decoding window},
cycle list={[samples of colormap={7 of customcolor}]},
legend columns=3 
	height=5.15cm,width=1.05\columnwidth,
	xmin=0.99, xmax=9.01,
        	ymin=7e-3,
	ymax=0.2,
	xtick distance=1,
	xlabel={Spatial position inside window},
	ylabel={BER},
	grid=both,
	ymode=log,
]
\addplot+[very thick,mark=none, mark options={}] table [x=POS, y=BER, col sep=comma] {data/Window32.csv};
\addlegendentry{$p_\text{win}=32$};
\addplot+[very thick,mark=square, mark options={}] table [x=POS, y=BER, col sep=comma] {data/Window33.csv};
\addlegendentry{$p_\text{win}=33$};
\addplot+[very thick,mark=diamond, mark options={}] table [x=POS, y=BER, col sep=comma] {data/Window34.csv};
\addlegendentry{$p_\text{win}=34$};
\addplot+[very thick,mark=*, mark options={}] table [x=POS, y=BER, col sep=comma] {data/Window35.csv};
\addlegendentry{$p_\text{win}=35$};
\addplot+[very thick,mark=triangle mark options={}] table [x=POS, y=BER, col sep=comma] {data/Window36.csv};
\addlegendentry{$p_\text{win}=36$};
\addplot+[very thick,mark=halfcircle, mark options={}] table [x=POS, y=BER, col sep=comma] {data/Window37.csv};
\addlegendentry{$p_\text{win}=37$};
\addplot+[very thick,mark=star, mark options={}] table [x=POS, y=BER, col sep=comma] {data/Window38.csv};
\addlegendentry{$p_\text{win}=38$};
\end{axis}

\end{tikzpicture}}
	\vspace*{-0.8cm}
	\caption{\ac{BER} within a window decoder with fixed $I=3$ iterations of size $w=9$ for several spatial positions during a decoder stall. The decoder gets stuck at around position 37. }
	\label{fig:ber-window}	
	\vspace*{-0.4cm}	
\end{figure}

\begin{figure*}
\begin{tabular}{ccc}
	\usetikzlibrary{matrix}
\usetikzlibrary{positioning}

\tikzstyle{box} = [draw,rounded corners=.1cm,inner sep=5pt,minimum height=5.75em, text width=7.5em, align=center,very thick] 

\begin{tikzpicture}
\def\circledarrow#1#2#3{ 
\draw[#1,->] (#2) +(80:#3) arc(80:-260:#3);
}  
\footnotesize
  \matrix (m) [matrix of nodes, row sep=2\pgflinewidth, column sep=2\pgflinewidth,
               nodes={rectangle, 
               minimum height=2em, minimum width=2em,
                      anchor=center, 
                      inner sep=0pt, outer sep=0pt},
                      left delimiter=.,
                      right delimiter=.,
                      ]
  {
     $\ddots$ & $\ddots$   \\
    &$\Hm_1$ & $\Hm_0$ \\
    && $\Hm_1$ & $\Hm_0$ \\
    && & $\Hm_1$ & $\Hm_0$  \\
    && & & $\Hm_1$ & $\Hm_0$  \\
    & && & & $\Hm_1$ & $\Hm_0$  \\
    && & & & &  $\ddots$ & $\ddots$   \\
  } ;
  
 \node [color=apfelgruen,box](w11)  at (-0.65,0.6){};
 \node[color=apfelgruen,above right= -0.5cm and -0.95cm of  w11] (t11){$t = 0$};
 
 \node [color=mittelblau,box](w12)  at (0.0,-0.0){};
  \node[color=mittelblau,above right= -0.5cm and -0.95cm of  w12] (t12){$t = 1$};
 
  \node [color=rot,box](w13)  at (0.65,-0.6){};
  \node[color=rot,above right= -0.5cm and -0.95cm of  w13] (t13){$t = 2$};
  

\draw[thick,->] (-2.5,-2.25) -- (2.5,-2.25) node[anchor=north east] {spatial pos.};

%
%
%
%

\end{tikzpicture}
&
	\usetikzlibrary{matrix}
\usetikzlibrary{positioning}

\tikzstyle{box} = [draw,rounded corners=.1cm,inner sep=5pt,minimum height=5.75em, text width=7.5em, align=center,very thick] 

\begin{tikzpicture}
\def\circledarrow#1#2#3{ 
\draw[#1,->] (#2) +(80:#3) arc(-90:200:#3);
}  
\footnotesize
  \matrix (m) [matrix of nodes, row sep=2\pgflinewidth, column sep=2\pgflinewidth,
               nodes={rectangle, 
               minimum height=2em, minimum width=2em,
                      anchor=center, 
                      inner sep=0pt, outer sep=0pt},
                      left delimiter=.,
                      right delimiter=.,
                      ]
  {
     $\ddots$ & $\ddots$   \\
    &$\Hm_1$ & $\Hm_0$ \\
    && $\Hm_1$ & $\Hm_0$ \\
    && & $\Hm_1$ & $\Hm_0$  \\
    && & & $\Hm_1$ & $\Hm_0$  \\
    & && & & $\Hm_1$ & $\Hm_0$  \\
    && & & & &  $\ddots$ & $\ddots$   \\
  } ;
  
 \node [color=apfelgruen,box](w11)  at (-0.65,0.6){};
 \node[color=apfelgruen,above right= -0.5cm and -0.95cm of  w11] (t11){$t = 0$};
 
 \node [color=mittelblau,box](w12)  at (0.0,-0.0){};
 \node [color=mittelblau,box,dashed](w12b)  at (0.075,-0.075){};
 \node[color=mittelblau,above right= -0.45cm and -0.95cm of  w12] (t12){$t = 1$};
 \node[color=mittelblau,above right= -0.7cm and -0.95cm of  w12] (t12b){$t = 2$};
 
 \node[color=mittelblau,above right= 0.05cm and -0.1cm of  w12] (text) {inc. $I$};
 \node (t1) at (1.5,0.2){};
 \circledarrow{thick, mittelblau}{t1}{0.45cm};
 
  \node [color=rot,box](w13)  at (0.65,-0.6){};
  \node[color=rot,above right= -0.5cm and -0.95cm of  w13] (t13){$t = 3$};
  

\draw[thick,->] (-2.5,-2.25) -- (2.5,-2.25) node[anchor=north east] {spatial pos.};

%
%
%
%
%

\end{tikzpicture}
&
	\usetikzlibrary{matrix}
\usetikzlibrary{positioning}

\tikzstyle{box} = [draw,rounded corners=.1cm,inner sep=5pt,minimum height=5.75em, text width=7.5em, align=center,very thick] 

\begin{tikzpicture}
\def\circledarrow#1#2#3{ 
\draw[#1,->] (#2) +(80:#3) arc(-20:120:#3);
}  

\def\circledarrowb#1#2#3{ 
\draw[#1,->] (#2) +(80:#3) arc(-180:-90:#3);
}

\footnotesize
  \matrix (m) [matrix of nodes, row sep=2\pgflinewidth, column sep=2\pgflinewidth,
               nodes={rectangle, 
               minimum height=2em, minimum width=2em,
                      anchor=center, 
                      inner sep=0pt, outer sep=0pt},
                      left delimiter=.,
                      right delimiter=.,
                      ]
  {
     $\ddots$ & $\ddots$   \\
    &$\Hm_1$ & $\Hm_0$ \\
    && $\Hm_1$ & $\Hm_0$ \\
    && & $\Hm_1$ & $\Hm_0$  \\
    && & & $\Hm_1$ & $\Hm_0$  \\
    & && & & $\Hm_1$ & $\Hm_0$  \\
    && & & & &  $\ddots$ & $\ddots$   \\
  } ;
  
 \node [color=mittelblau,box](w11)  at (-0.65,0.6){};
 \node[color=mittelblau,above right= -0.5cm and -0.95cm of  w11] (t11){$t = 1$};
 
  \node[color=mittelblau,above right= -1.65cm and -3.12cm of  w11] (text2) {shift forward};
 \node (t2) at (-2.1,0.2){};
 \circledarrowb{thick, mittelblau}{t2}{1.2cm};
 
 \node [color=apfelgruen,box](w12)  at (0.0,-0.0){};
 \node[color=apfelgruen,above right= -0.45cm and -0.95cm of  w12] (t12){$t = 0$};

 \node[color=apfelgruen,above right= 0.05cm and -0.55cm of  w12] (text) {shift back};
 \node (t1) at (1.25,0.4){};
 \circledarrow{thick, apfelgruen}{t1}{0.45cm};
 
  \node [color=rot,box](w13)  at (0.65,-0.6){};
  \node[color=rot,above right= -0.5cm and -0.95cm of  w13] (t13){$t = 2$};


\draw[thick,->] (-2.5,-2.25) -- (2.5,-2.25) node[anchor=north east] {spatial pos.};

%
%
%
%
%

\end{tikzpicture}\\
\footnotesize{(a) Conventional Windowed Decoder} &\footnotesize{(b) Adaptive Iterations Decoder} & \footnotesize{ (c) Window Shift Decoder}\\
\end{tabular}
\vspace*{-0.1cm}
\caption{Different window decoding schemes.}
\label{fig:decoder-visualization}		
\vspace*{-0.5cm}
\end{figure*}
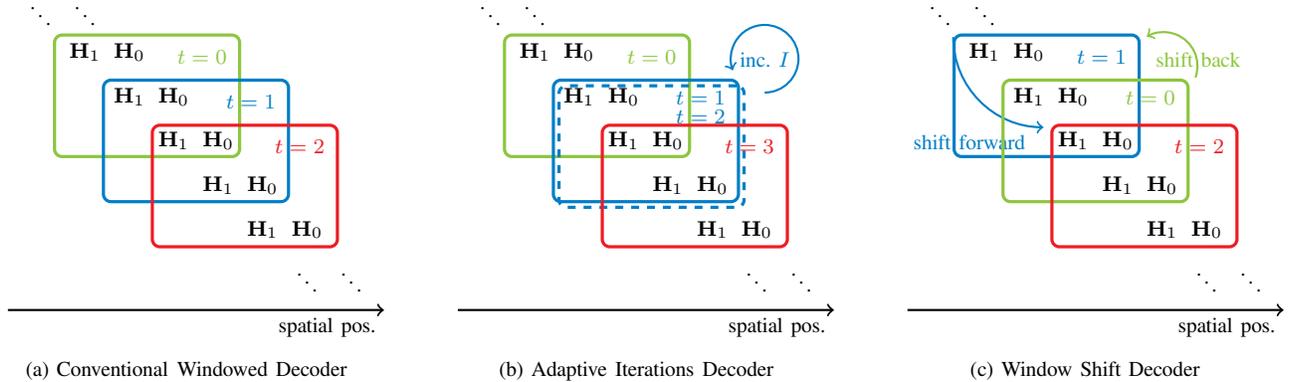

Contrary to full BP decoding, windowed decoding may suffer from an effect called \emph{decoder stall} (see \cite{Schmalen2016Window} and similarly \cite{zhu2018braided}) caused by unrecoverable errors within the active part of the decoding window. Due to this effect, the decoding performance decreases significantly as subsequent sub-blocks rely on these erroneous blocks. Thus, the observed effect at the decoder output is similar to burst-like error patterns. We define the stall position $s$ as the first erroneous block after decoding, i.e.,
$$s=\arg\min_{i} \left( \sum_{j=i}^{i+\eta-1}\mathbbm{1}_{\{{P}_{\text{e,post},j} >\delta\}}=\eta \right)$$
where $P_{\text{e,post},j}$ denotes the \ac{BER} after decoding at spatial position $j$ and $\delta$ is a fixed threshold to exclude the case of error floor patterns, e.g., due to absorbing sets. Further, $\eta$ defines the amount of consecutive erroneous blocks until a stall is marked  (e.g., $\eta=\mu$) and $\mathbbm{1}_{\{x>\delta\}}$ denotes the indicator function, i.e., returns 1 if $x>\delta$ and 0 otherwise.
Whenever this stall occurs at start position $s$, all subsequent $L-s$ blocks are typically erroneous without further resynchronization strategy (e.g., as proposed in \cite{zhu2018braided}). Using \cite{Schmalen2016Window}
\begin{equation}
P_{\text{e,post}} \approx \biggl(1-\frac{\mathbb{E}[s]}{L}\biggr) \cdot P_{\text{e,pre}}\cdot P_{\text{stall}}.    
\label{equ:BER}
\end{equation}
we can approximate the BER after decoding, influenced by the decoder stall. The required parameters are the average start position of the decoder stall $\mathbb{E}[s]$, the average BER before decoding $P_{\text{e,pre}}$ (i.e., the \ac{BER} at the channel output) and the probability of occurrence of a decoder stall $P_{\text{stall}}$.   

It is important to mention that this problem occurs only in the region between the \emph{wave-threshold}  $E_b/N^*_{0,\text{coupled}}$ and the \emph{macro-threshold} $E_b/N^*_{0,\text{macro}}$ \cite{Schmalentenbrink}, i.e., the \emph{BP-threshold} of the underlying block code. For the code used throughout this work, asymptotically this region is between $E_b/N^*_{0,\text{coupled}}=2.117$ dB and $E_b/N^*_{0,\text{macro}}=2.896$ dB. Above this region, for sufficiently high SNR\footnote{Note that the thresholds are asymptotic thresholds and, thus, only provide first approximation in the finite length regime assuming infinite iterations and $n \to \infty$.} the decoder can recover.

The underlying effect for a windowed decoder with $I=3$ fixed iterations per window is shown in Fig.~\ref{fig:ber-window}. 
In this example the BER-distribution inside different windows $p_\text{win}$ around the stall start position is shown. The decoder gets stuck at around position $s=37$. First, it is important to mention that at least the left-most position inside the window has to be error-free as this equals the decoder output. As the performed $I=3$ iterations per window shift are not enough for successful decoding, the \ac{BER} of the first spatial position within the decoder starts to increase (i.e., the decoder ``\emph{loses track of the wave}'').
An increased number of $I=4$ iterations solves the problem, but also increases the average decoding complexity by $33\%$.


\subsection{Test-set creation and evaluation}

The occurrence of such decoder stalls is rare but contributes significantly to the \ac{BER} as many spatial positions are effected. As this effect does not show up in the \ac{DE} and an analytical prediction does not exist so far, we decided to evaluate different approaches based on a test-set of noisy codewords. This test-set is created using a windowed decoder implemented in CUDA that is capable of fast decoding using \acp{GPU}. Whenever the decoder stalls, the noisy codeword is saved in the test-set. 
The final test-set contains 2,000 codewords after passing an AWGN channel with $\textrm{SNR}_\text{b}$ = 2.86 dB. The windowed decoder uses a window size of $w = 9$ and $I$ = 3 iterations per window. For the test-set, we find $\mathbb{E}[s] = 57.79$ ($L=99$). The starting positions $s$ is approximately uniformly distributed for $s \in [10,90]$, with lower stall probability at the boundaries due to termination.



\section{Adaptive Window Decoding Schemes}

An intuitive explanation for $P_{\text{stall}}$ is the observation in Sec.~\ref{sec:intro_sc} that in a few (rare) cases, the number of performed iterations is not sufficient for this specific noise realization (see Fig.~\ref{fig:ber-window}, when using $I=4$ instead of $I=3$ fixed iterations, decoding is successful), i.e., the decoder needs a locally increased number of iterations.
Obviously, an increased fixed number of iterations solves the problem, but also increases the decoding complexity.  
In the following, we analyze different strategies to decrease $P_{\text{stall}}$ while keeping the overall decoding complexity low.
Note: we do not consider larger window sizes $w$ as this would inherently require more hardware units.

\subsection{Stall detection}

To allow adaptive windowed decoding, the decoder needs to track the decoding progress continuously. This can be done by two different approaches:
\begin{enumerate}
\item Parity-check-based: verify whether all parity-checks within a certain window-position are fulfilled or not.
\item Log likelihood ratio (LLR)-based: estimate the soft-\ac{BER} based on the message \acp{LLR} according to \cite{hoeher2000log} 
\begin{equation}
P_{e,\text{est}} = \frac{1}{K} \sum_{k=1}^{K} \frac{1}{1 + \exp({|L_k}|)}.
\label{equ:Paritysoft}
\end{equation}
\end{enumerate}

The second method uses an estimation of the \ac{BER} within each sub-block. Due to this fact there can be no hard decision rule, but a threshold $\delta_\text{BER}$ used for stall detection. 


In our experiments, we observe that a decision only based on the first spatial position within the decoder does not yield reliable decisions. Thus, we evaluate the decoding progress at third spatial position (keep in mind: $\mu=1$) leading to more robust decisions. An intuitive explanation can be given by the fact that typically the first positions within the decoder window are almost error-free (e.g., see $p_{win}=32$ in Fig. \ref{fig:ber-window}), while the third position is more sensitive to looming stalls.
Note that the method used for stall detection is independent of the following decoding algorithms.

\subsection{Adaptive iterations decoder (AID)} 
\begin{algorithm}[t]
	\renewcommand{\algorithmicrequire}{\textbf{Input:}}
	\renewcommand{\algorithmicensure}{\textbf{Output:}}
	\caption{Adaptive iterations decoder}
	\label{alg1}
	\begin{algorithmic}
		\Require			
		\State {$I_\text{min}$ \qquad min. number of iter. per window}
		\State{$I_\text{max}$ \qquad max. number of iter. per window}
		\For{$p_\text{win}=1:N_\text{win}$}
		\State{$I\gets 0$}
		\While{$I < I_\text{max}$}
		\State CN update, VN update and stall detection 
		\If{$I \geq I_\text{min}$ \textbf{and} stall detection == \textbf{false}}
		\State break
		\EndIf
		\State $I\gets I+1$
		\EndWhile
		\EndFor
	\end{algorithmic}
\end{algorithm}

The first adaptive decoder uses a trivial concept of increasing the number of iterations $I$ to prevent decoder stalls as described in Alg.~\ref{alg1} and Fig.~\ref{fig:decoder-visualization}. The value $I_\text{min}$ defines the minimum number of iterations per window that are performed in case there is no decoder stall detected. In the other case the decoder can carry out at most $I_\text{max}$ iterations per window to prevent from a decoder stall.

\subsection{Window shift decoder (WSD)}
The second adaptive decoder uses a different approach by shifting the window backwards as described in Alg.~\ref{alg2} and Fig.~\ref{fig:decoder-visualization}. The number of iterations per window and the window size are constant. The parameters $I_\text{min}$ and $I_\text{max}$ follow the same definition as in Alg.~\ref{alg1} while another parameter $n_\text{b}$ is introduced that denotes the number of positions the window is shifted backwards in case of an emerging decoder stall.

\begin{algorithm}[t]
	\renewcommand{\algorithmicrequire}{\textbf{Input:}}
	\renewcommand{\algorithmicensure}{\textbf{Output:}}
	\caption{Window shift decoder}
	\label{alg2}
	\begin{algorithmic}
		\Require			
		\State {$I_\text{min}$ \qquad min. number of iter. per window}
		\State{$I_\text{max}$ \qquad max. number of iter. per window}
		\State{$n_{b}$   \qquad \quad number of skipped blocks}
		\For{$p_\text{win}=1:N_\text{win}$}
		\State{$I\gets 0$}
		\While{$I < I_\text{max}$}
		\State CN update, VN update and stall detection 
		\If{$I$ == $I_\text{min}$ \textbf{and} stall detection == \textbf{false}}
		\State break
		\ElsIf {$I$ == $I_\text{min}$}
		\State $P_{\text{stall}}\gets p_{\text{win}} - n_\text{b}$  
		\EndIf
		\State $I\gets I+1$
		\EndWhile
		\EndFor
	\end{algorithmic}

\end{algorithm}

\subsection{Wave tracking decoder (WTD)}
The third adaptive decoder concept is an enhanced version of the window shift decoder. The intuition behind this decoder is to shift the window according to the position of the decoding wave as explained in Alg.~\ref{alg3} and Fig.~\ref{fig:decoder-visualization}. Thus, this decoder can shift the decoding window backwards and forward based on the position of the decoding wave. Therefore, the decoding wave is mostly kept on the left inside the decoding window and unnecessary updates of error-free positions are minimized. The input parameters of the decoder are the same as in Alg.~\ref{alg2}.      

\begin{algorithm}[t]
	\renewcommand{\algorithmicrequire}{\textbf{Input:}}
	\renewcommand{\algorithmicensure}{\textbf{Output:}}
	\caption{Wave tracking decoder}
	\label{alg3}
	\begin{algorithmic}		
		\Require			
		\State {$I_\text{min}$ \qquad min. number of iter. per window}
		\State{$I_\text{max}$ \qquad max. number of iter. per window}
		\State{$n_{b}$   \qquad \quad number of skipped blocks}
		\For{$p_\text{win}=1:N_\text{win}$}
		\State flag $\gets 0$
		\State{$I\gets 0$}
		\While{$I < I_\text{max}$}
		\State CN update, VN update and stall detection 
		\If{$I$ == $I_\text{min}$ \textbf{and} stall detection == \textbf{false}}
		\State break
		\ElsIf {$I$ == $I_\text{min}$}
		\State $p_{\text{win}}\gets p_{\text{win}} - n_\text{b}$
		\State flag=1
		\EndIf
		\If {$I > I_\text{min}$ \textbf{and} stall detection == \textbf{false} \textbf{and} flag==1 }
		\State $p_{\text{win}}\gets p_{\text{win}} + n_\text{b}$ 
		\State flag=0
		\ElsIf {$I > I_\text{min}$ \textbf{and} stall detection == \textbf{false}}
		\State $p_{\text{win}}\gets p_{\text{win}}+1$ 
		\EndIf
		\State $I\gets I+1$
		\EndWhile
		\EndFor
	\end{algorithmic}
\end{algorithm}

\section{Stall Prediction vs. Stall Detection}

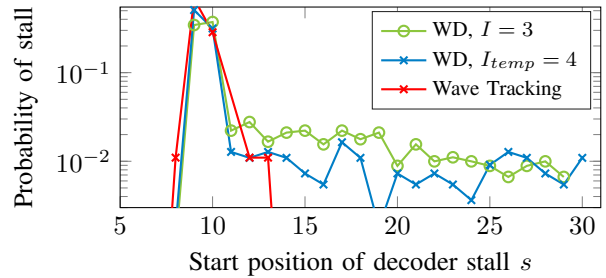
\begin{figure}[t]
	\vspace*{-0.3cm}
	\centering
	\begin{tikzpicture}
\begin{axis}[
height=4.25cm,width=0.9\columnwidth,
ymin=0, ymax=0.55,ymin=3e-3,
ymode=log,
xmin=5, xmax=31,
legend style={legend cell align=left,align=left,draw=white!15!black, font=\footnotesize},
xlabel={Start position of decoder stall $s$},
ylabel={Probability of stall},
]

\addplot[thick,mark=o, apfelgruen] table [x=POS, y=BER, col sep=comma]
{./data/Histogram1.1dbstatic3.csv};
\addlegendentry{WD, $I = 3$}

\addplot[thick,mark=x, mittelblau] table [x=POS, y=BER, col sep=comma]
{./data/Histogram1.1dbtemp4s8e12.csv};
\addlegendentry{WD, $I_{temp}= 4$}

\addplot[thick,mark=x, red] table [x=POS, y=BER, col sep=comma]
{./data/Histogram1.1dbWave.csv};
\addlegendentry{Wave Tracking}

\end{axis}
\end{tikzpicture}
		\vspace*{-0.3cm}
	\caption{Distribution of decoder stalls of different decoders for manipulated block 10 with $\textrm{SNR}_\text{b,manip} = 2.1\,dB$, $\textrm{SNR}_\text{b} = 2.86\,dB$, $I = 3$ and $w= 9$.}
	\label{pgf:StallHistogram}		
	\vspace*{-0.7cm}
\end{figure}

So far, all proposed decoders rely on a stall detection, i.e., the decoder \emph{acts} whenever a stall is likely in progress.
However, it may require more iterations to recover from such a state in contrast to a decoder which predicts potential issues already based on its input channel \acp{LLR}. 
To answer this question, we conduct the following experiment. 
We manually create codewords that lead to decoder stalls. This is achieved by manipulation of the \ac{SNR} within single spatial position of the code word (the position is known during this experiment). The parameters of this experiment are the same as used for test-set creation, however, we only use noisy codewords (i.e., a specific noise realization) which can be successfully decoded with the naive windowed decoder. Now, we replace block 10 (to avoid influences of the boundary) by adding noise such that block 10 has a reduced $\textrm{SNR}_\text{b,manip}$, i.e., we add artificial noise. Afterwards, the decoder has to decode the manipulated codewords. This procedure guarantees that the decoder stall is caused by the corrupted block at spatial position 10.

In Fig. \ref{pgf:StallHistogram}, the distribution of decoder stalls is visualized for different decoders. 
In the first case, using constant $I$ = 3, we can notice that only approximately 18\% of the manipulated codewords cause the decoder to stall (strongly depending on $\textrm{SNR}_\text{b,manip}$ of course). The more interesting observation is the distribution of the decoder stall start position $s$. The average value amounts to $\mathbb{E}[s]$ = 11.1. This is something we could have expected. 
But somehow to our surprise, the corrupted block also causes decoder stalls that are even more than a complete decoding window length away.

Now, we use the fact that we know which position causes the decoder to stall and check if this can reduce the frequency of stall occurrence. Therefore, we temporarily increase the number of iterations to $I$ = 4 starting at position $L_\text{s} = 8$ until $L_\text{e} = 12$. Thereby, the frequency of stall occurrence can be reduced by approximately 40\%, but the \emph{tail-like} distribution of $s$ is still present. Finally, the wave tracking decoder (stall detection) does not show such a behavior.
This leads to the conclusion that predicting stalls is not necessarily leading to successful prevention of decoder stalls. The reason is that even if we know which position causes the decoder stall, we can state that it is more likely to stall next to the predicted position, but due to the distribution of $s$ also at subsequent positions.  
With a pure prediction, the decoder would have to increase the number of iterations to $I=5$ whenever a stall may occur. However, in many cases $I=4$ is simply sufficient and, thus, the average complexity of a predictive decoder is larger than its retrospective counterpart.\footnote{Remark: Obviously a combination of both methods may further improve the performance.}

\section{BER Performance and Complexity}

\begin{table}
	\centering
	\caption{Average complexity of different windowed decoders for $\textrm{SNR}_\textrm{b} = 2.86\,dB$ and fixed $w=9$}
		\vspace*{-0.2cm}
	\begin{tabular}{ll}
		Decoder & Avg. Complexity \\
		\hline
		WD, $I = 3$ & 3 $\cdot w$\\
		WD, $I = 4$ & 4 $\cdot w$ \\
		Adaptive Iterations  & 3.01 $\cdot w$  \\
		Window Shift  & 2.77 $\cdot w$  \\
		Wave Tracking  & \textbf{2.64} $\cdot w$  \\
		\hline
	\end{tabular}
	\label{tab:Complexity}
	\vspace*{-0.5cm}
\end{table}

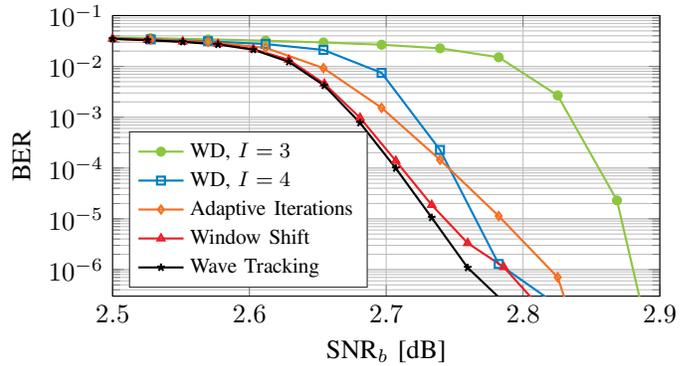
\begin{figure}
	\centering
	\begin{tikzpicture}

\pgfplotsset{compat=1.12}
\begin{axis}[
        width=\linewidth,
	    height=0.6\linewidth,
        xmajorgrids,
        yminorticks=true,
        ymajorgrids,
        yminorgrids,
        legend pos=south west,        
        legend style={legend cell align=left,align=left,draw=white!15!black, font=\footnotesize},
        xlabel={$\textrm{SNR}_b$ [dB]},
        ylabel={BER},
        ymode=log,
        mark size=1.5pt,
        xmin=2.5,
        xmax=2.9,
        ymin=3e-7,
        ymax=0.1
    ]	

	\addplot[mark=*,color=apfelgruen, thick] table [x expr=\thisrowno{0}+1] {data/results_static3.dat};
	\addlegendentry{WD, $I = 3$}

	\addplot[mark=square,color=mittelblau, thick] table [x expr=\thisrowno{0}+1] {data/results_static4.dat};
	\addlegendentry{WD, $I = 4$}
	
	\addplot[mark=diamond,color=orange, thick] table [x expr=\thisrowno{0}+1] {data/resultsmin35.dat};
	\addlegendentry{Adaptive Iterations}
	
	\addplot[mark=triangle,color=rot, thick] table [x expr=\thisrowno{0}+1] {data/results_3back.dat};
	\addlegendentry{Window Shift}
	
	\addplot[mark=star,color=black, thick] table [x expr=\thisrowno{0}+1] {data/results_4andback.dat};
	\addlegendentry{Wave Tracking}

\end{axis}

\end{tikzpicture}
	\vspace*{-0.8cm}
	\caption{BER performance for different decoders.}
	\label{pgf:BER}		
	\vspace*{-0.5cm}
\end{figure}

The \ac{BER} performance of all proposed decoders is shown in Fig.~\ref{pgf:BER}. Further, Table~\ref{tab:Complexity} shows the average complexity for the different decoding schemes. All decoders use the \ac{LLR}-based stall detection with $\delta_\text{BER}=10^{-7}$.
The fixed iteration decoders either suffer from poor \ac{BER} performance or significantly higher complexity of approximately $33\%$. The adaptive iteration decoder improves both quantities, however, a simple adaptive iteration windowed decoder does not provide the best trade-off.
Better results are achieved by a window shift.
The intuition behind can be given by looking at the \emph{decoding wave} within the decoder (see Fig.~\ref{fig:ber-window}): the largest \ac{LLR} updates are expected in the \emph{center} of the wave. As the iterative decoder increases the number of iterations whenever the \ac{BER} decreases in the first positions (i.e., the decoder starts losing track of the wave) the \ac{LLR} updates are, on average, less effective. Thus, the decoder is conceptually disadvantaged when compared to a window shift decoder. Finally, the Wave Tracking approach combines the best \ac{BER} performance with the (almost) lowest decoding complexity.


\section{Conclusions and outlook}\label{sec:conclusions}

In this work, we have analyzed different adaptive windowed decoding strategies and their influence on the decoding performance. It turned out that an adaptive window shift performs significantly better, i.e., yields the same $P_\text{stall}$ while having lower overall complexity. 
We have shown by experiment that a foresightful stall prediction does not necessarily outperform an retrospective stall prediction, which allows easier implementations. 
As a result, the average decoding complexity of the Wave Tracking decoder could be decreased by $34\%$ when compared to a windowed decoder with four fixed iterations.

\bibliographystyle{IEEEtran}
\bibliography{IEEEabrv,references}

\end{document}